\documentclass[jcp,reprint,superscriptaddress]{revtex4-1}

\usepackage{amsmath, bm, amssymb}
\usepackage{graphicx}
\usepackage{here}

\begin{document}

\title{An Efficient Matrix Product Operator Representation of the Quantum-Chemical Hamiltonian}

\author{Sebastian Keller}
\email{sebastian.keller@phys.chem.ethz.ch}
\affiliation{ETH Z\"urich, 
Laboratory of Physical Chemistry, 
Vladimir-Prelog-Weg 2, 8093 Z\"urich, Switzerland}

\author{Michele Dolfi}
\email{dolfim@phys.ethz.ch}
\affiliation{ETH Z\"urich,
Institute of Theoretical Physics
Wolfgang-Pauli-Strasse 27, 8093 Z\"urich, Switzerland}

\author{Matthias Troyer}
\email{troyer@itp.phys.ethz.ch}
\affiliation{ETH Z\"urich,
Institute of Theoretical Physics
Wolfgang-Pauli-Strasse 27, 8093 Z\"urich, Switzerland}

\author{Markus Reiher}
\email{markus.reiher@phys.chem.ethz.ch}
\affiliation{ETH Z\"urich, 
Laboratory of Physical Chemistry, 
Vladimir-Prelog-Weg 2, 8093 Z\"urich, Switzerland}

\begin{abstract}
We describe how to efficiently construct the quantum chemical Hamiltonian operator in matrix product form.
We present its implementation as a density matrix renormalization group
(DMRG) algorithm for quantum chemical applications in a purely matrix product based framework.
Existing implementations of DMRG for quantum chemistry
are based on the traditional formulation of the method, which was developed from a viewpoint of Hilbert space decimation
and attained a higher performance compared to straightforward implementations of
matrix product based DMRG.
The latter variationally optimizes a class of ansatz states known as matrix product states (MPS),
where operators are correspondingly represented as matrix product operators (MPO).
The MPO construction scheme presented here eliminates the previous performance disadvantages
while retaining the additional flexibility provided by a matrix product approach;
for example, the specification of expectation values becomes an input parameter.
In this way, MPOs for different symmetries  --- abelian and non-abelian  --- and different relativistic and
non-relativistic models may be solved by an otherwise unmodified program.
\end{abstract}

\maketitle
%\keywords{DMRG, Matrix Product Operators}

%%%%%%%%%%%%%%%%%%%%%%%%%%%%%%%%%%%%%%%%%%%%%%%%%%%%%%%%%%%%%%%%%%%%%%%%%%%%%%%%
\section{Introduction}
\label{sec:intro}
%%%%%%%%%%%%%%%%%%%%%%%%%%%%%%%%%%%%%%%%%%%%%%%%%%%%%%%%%%%%%%%%%%%%%%%%%%%%%%%%

The density matrix renormalization group (DMRG) algorithm
proposed by White \cite{White1992, White1993} to solve one-dimensional 
effective models in condensed matter theory has been established
as a powerful method to tackle strongly correlated systems in quantum chemistry
\cite{Marti2010a, Legeza2008, Chan2011a, Wouters2014review, Kurashige2014, Yanai2015, Legeza2015, Marti2011}.
In contrast to other active space methods such as CASSCF and CASCI, much larger
active spaces can be treated.

The original DMRG algorithm formulated by White is a variant of other renormalization group
methods, relying on Hilbert space decimation and reduced basis transformations.
Instead of truncating the eigenstates of the Hamiltonian according to their energy, the selection
of eigenstates based on their weight in reduced density matrices dramatically improved
the performance for one-dimensional systems studied in condensed matter physics.

An important contribution towards the present understanding of the algorithm was made by
\"Ostlund and Rommer \cite{Oestlund1995}, who showed that the block states in DMRG can be
represented as matrix product states (MPS). They permitted predictions \cite{Vidal2003, Barthel2006, Schuch2008} for the decay of the spectra
of reduced density matrices via area laws \cite{Bekenstein1973, Srednicki1993} for the entanglement entropy and by
Verstraete et al.\ \cite{Verstraete2004b}, who derived the DMRG algorithm from a variational principle.

At first, MPS were mainly a formal tool to study the properties of DMRG; an application for numerical purposes
was only considered for special cases, such as calculations under periodic boundary conditions \cite{Verstraete2004b}.
The usefulness of the matrix-product formalism for general applications was demonstrated by
McCulloch \cite{McCulloch2007} and by Crosswhite et al.\ \cite{Crosswhite2008} who
employed the concept of matrix product operators (MPO) to represent Hamiltonian operators.
MPOs were previously introduced in Ref.\ \cite{Verstraete2004} to calculate finite-temperature density matrices.

The main advantage of MPS is that they encode wave functions as stand-alone objects that can directly be
manipulated arithmetically as a whole.
In traditional DMRG by contrast, a series of reduced basis transformations entail a sequential dependency
dictating when a certain part of the stored information becomes accessible.
An MPS-based algorithm may therefore process the information from independently computed wave functions.
McCulloch \cite{McCulloch2007} exploited this fact to calculate excited states
by orthogonalizing the solution of each local DMRG update against previously calculated lower lying states.
This state-specific procedure converges excited states fast and avoids the performance penalty incured by
a state-average approach, where all states must be represented in one common and consequently larger basis.
%
%In traditional DMRG, excited state calculations require a state-average treatment in which the
%ground and excited states are computed in a single calculation and
%renormalized block states $m$ (see Sec.\ \ref{sec:mps}) are divided among the ground and excited states.
%As all variants of DMRG scale as $\mathcal{O}(m^3)$, a state-average approach incurs a high performance
%penalty, while the computational cost per excited state stays constant for an MPS algorithm.
%
We are not aware of any excited-state algorithm based on projecting out lower states
in the context of traditional DMRG,
%excited state calculations by orthogonalizing against lower lying states with quantum chemical DMRG were performed by Wouters et al.\ \cite{Wouters2014}
but note that Wouters et al.\ \cite{Wouters2014} employed such an algorithm
in the \textsc{CheMPS2} DMRG program, which respresents the wave function as an MPS, but employs a traditional operator format.
We may refer to these traditional DMRG programs in quantum chemistry as first-generation programs and denote a truly MPO-based 
implementation of the DMRG algorithm a second-generation DMRG algorithm in quantum chemistry. Such a second-generation 
formulation allows for more flexibility in finding the optimal state and allows for a wider range of applications.
The possibility to store the wave function in the MPS form allows us to perform (complex) measurements at a later time without the need to 
perform a full sweep through the system for the measurement.

In the MPO formalism, we can perform operator arithmetic, such as additions and multiplications.
An example where this is useful is the calculation of the variance $\Delta E = \langle H^2 \rangle - \langle H \rangle^2$, requiring
the expectation value of the squared Hamiltonian.
Since the variance vanishes for an exact eigenstate, it is a valuable measure to assess the quality of DMRG wave functions, albeit
of limited relevance for quantum chemistry, because quantum chemical DMRG calculations are limited by the size of the Hamiltonian MPO,
whose square is too expensive to evaluate for large systems.
%
%While in traditional DMRG, the action of each term of the squared operator
%would have to be implemented manually, an operator in the MPO format can be squared directly as described in Ref. \cite{McCulloch2007}.
%Variance calculations with squared MPOs are most practicable for the effective models studied by the condensed matter physics community
%with simple Hamiltonians, such as the Fermi-Hubbard model \cite{Hubbard1963}, because the resulting MPOs are still numerically feasible.
%The square of the quantum chemical Hamiltonian MPO in contrast, is computationally very expensive to evaluate due to the large number of terms
%and is consequently only accessible for small systems.
%
But apart from the possibility of operator arithmetic, the adoption of MPOs for quantum chemical DMRG has additional advantages.
The MPO structure inherently decouples the operator from core program routines performing operations on MPS and MPOs.
The increased flexibility for Hamiltonian operators, on the one hand,
permitted us to implement both abelian and non-abelian symmetries 
with only one set of common contraction routines. We note, that the relativistic Coulomb-Dirac Hamiltonian can also be solved
by the same set of program routines, which we will describe in a later work.
Measuring observables, on the other hand, requires no additional effort;
only the locations of the elementary creation and annihilation operators need to be specified
for a machinery capable of calculating expectation values of arbitrary MPSs and MPOs.
The 26 different correlation functions required to calculate single- and two-site entropies \cite{Rissler2006, Boguslawski2013b}, for example,
can thus be specified as a program input.

The algorithm that we develop in this work is implemented based on the ALPS MPS program \cite{Dolfi2014},
which is a state-of-the-art MPO-based computer program applied in the condensed matter physics community.
To its quantum chemical version presented in this work we refer to as \textsc{QCMaquis}.

This work is organized as follows. In Secs.\ \ref{sec:mps} and \ref{sec:mpo} we introduce the basic concepts of MPS and MPO respectively.
The inclusion of fermionic anticommutation is discussed in Sec.\ \ref{sec:fermions}. In Sec.\ \ref{sec:eval}, we describe how
MPOs act on MPSs. Finally, we apply our implementation to the problem of two-dimensionally correlated electrons in graphene fragments in Sec.\ \ref{sec:calculations}.

%%%%%%%%%%%%%%%%%%%%%%%%%%%%%%%%%%%%%%%%%%%%%%%%%%%%%%%%%%%%%%%%%%%%%%%%%%%%%%%%
\section{Matrix Product States}
\label{sec:mps}
%%%%%%%%%%%%%%%%%%%%%%%%%%%%%%%%%%%%%%%%%%%%%%%%%%%%%%%%%%%%%%%%%%%%%%%%%%%%%%%%

Any state $|\psi\rangle$ in a Hilbert space spanned by $L$ spatial orbitals
can be expressed in terms of its configuration interaction (CI) coefficients $c_{\bm{\sigma}}$
with respect to occupation number vectors $|\bm{\sigma}\rangle$.
\begin{equation}
    | \psi \rangle = \sum_{\bm{\sigma}} c_{\bm{\sigma}} |\bm{\sigma} \rangle,
\label{eq:state}
\end{equation}
with
 $|\bm{\sigma} \rangle =
|\sigma_1, \, \ldots, \, \sigma_L \rangle$, and
$\sigma_l = |\!\! \uparrow \! \downarrow \rangle , | \! \uparrow \, \rangle , | \! \downarrow \rangle , | \,0\, \rangle$.
For MPS, the CI coefficients $c_{\bm{\sigma}}$ are encoded
as the product of $L$ matrices $M^{\sigma_i}$, explicitly written as
\begin{equation}
    c_{\bm{\sigma}} = 
        \sum_{a_1, \ldots, a_{L-1}} \!\!\!\!\!
        M^{\sigma_1}_{1 a_1} \, M^{\sigma_2}_{a_1 a_2} \, \cdots \, M^{\sigma_L}_{a_{L-1} 1}
        \,\,\,\,
        ,
    \label{eq:mps_coeff}
\end{equation}
so that the quantum state reads
\begin{equation}
| \psi \rangle = \sum_{\bm{\sigma}}
        \sum_{a_1, \ldots, a_{L-1}} \!\!\!\!\!
        M^{\sigma_1}_{1 a_1} \, M^{\sigma_2}_{a_1 a_2} \, \cdots \, M^{\sigma_L}_{a_{L-1} 1}
        \,\,
        |\bm{\sigma} \rangle.
    \label{eq:mps_long}
\end{equation}
Collapsing the explicit summation over the $a_i$ indices in the previous equation as matrix-matrix
multiplications, we obtain in compact form
\begin{equation}
| \psi \rangle = \sum_{\bm{\sigma}}
        M^{\sigma_1} \, M^{\sigma_2} \, \cdots \, M^{\sigma_L}
        |\bm{\sigma} \rangle,
    \label{eq:mps}
\end{equation}
where $M^{\sigma_1}$, $M^{\sigma_l}$ and $M^{\sigma_L}$ ($l<1<L$)  are required to have
matrix dimensions $ 1 \times m_1 $, $ m_{l-1} \times m_{l}$, and $ m_{L-1} \times 1 $, respectively,
as the above product of matrices must yield the scalar coefficient $c_{\bm{\sigma}}$.

The reason why Eq.\ \eqref{eq:mps} is not simply a more complicated way of writing
the CI expansion in Eq.\ \eqref{eq:state} is that the dimension of the matrices $M^{\sigma_l}$
may be limited to some maximum dimension $m$, referred to as the number of renormalized block states \cite{White1993}.
This restriction is the key idea that reduces the exponential growth of the original Full-CI (FCI) expansion
to a polynomially scaling ansatz wave function.
While the evaluation of all CI coefficients $c_{\bm{\sigma}}$ is still exponentially expensive,
the ansatz in Eq.\ \eqref{eq:mps} allows one to compute
scalar products $\langle \psi| \psi \rangle$ and expectation values $\langle \psi | \widehat{O} | \psi \rangle$ of an operator
$\widehat{O}$ in polynomial time as shown in section\ \ref{sec:expectation_values}.
At the example of the norm calculation
\begin{align}
    \langle \psi | \psi \rangle &= \sum_{\bm{\sigma}} c^*_{\bm{\sigma}} c_{\bm{\sigma}}\\  \nonumber
                                &= \sum_{\bm{\sigma}} M^{\sigma_L \dagger} \cdots M^{\sigma_1 \dagger} M^{\sigma_1} \cdots M^{\sigma_L},
\end{align}
we observe that the operational complexity only reduces from exponential to polynomial,
if we group the summations as
\begin{equation}
  \langle \psi|\psi \rangle = \sum_{\sigma_L} M^{\sigma_L \dagger} \cdots \sum_{\sigma_1} \Big(M^{\sigma_1 \dagger} M^{\sigma_1} \Big) \cdots M^{\sigma_L},
\end{equation}
in addition to limiting the matrix sizes.

The price to pay, however, is that such an omission of degrees of freedom leads to the introduction
of interdependencies among the coefficients. Fortunately, in molecular systems,
even MPS with a drastically limited number of $m$ accurately
approximate ground and low-lying excited state energies (see Refs.\ \cite{Marti2008, Kurashige2011, Sharma2012} for examples).
Moreover, MPS are systematically improvable towards the FCI limit by increasing $m$ and extrapolation techniques are available
to estimate the distance to this limit for a simulation with finite $m$.

Despite the interesting properties of MPS, Eq.\ \eqref{eq:mps} might still not
seem useful, as we have not discussed how the $M^{\sigma_l}$
matrices can be constructed for some state
in the form of Eq.\ \eqref{eq:state}.
Although the transformation from CI coefficients to the MPS format is possible \cite{Schollwock2011},
it is of no practical relevance,
because the resulting state would only be numerically manageable
for small systems, where FCI would be feasible as well\ \cite{Moritz2007}.
DMRG calculations may, for instance, be started from a random\ \cite{Chan2004}
or from a state guess based on quantum informational criteria\ \cite{Legeza2003, Legeza2004}.
The lack of an efficient conversion to and from a CI expansion is therefore not a disadvantage
of the MPS ansatz. We also note that given an MPS state, the most important CI coefficients can be reconstructed even
for huge $\lbrace|\bm{\sigma}\rangle \rbrace$ spaces\ \cite{Boguslawski2011a}.

MPS that describe eigenstates of some given Hamiltonian operator inherit the symmetries of the latter and the constitutent
MPS tensors $M^{\sigma_i}$ therefore feature a block diagonal structure. Examples of how block diagonal tensors can be exploited
in numerical computations may be found in Refs. \cite{Bauer2011} and \cite{Dolfi2014}.
For additional information about MPS such as left and right normalization with
singular value or QR decomposition we refer to Ref.\ \cite{McCulloch2007} and to the detailed review
by Schollw\"ock\ \cite{Schollwock2011}.

%%%%%%%%%%%%%%%%%%%%%%%%%%%%%%%%%%%%%%%%%%%%%%%%%%%%%%%%%%%%%%%%%%%%%%%%%%%%%%%%
\section{Matrix Product Operators}
\label{sec:mpo}
%%%%%%%%%%%%%%%%%%%%%%%%%%%%%%%%%%%%%%%%%%%%%%%%%%%%%%%%%%%%%%%%%%%%%%%%%%%%%%%%

Formally, the MPS concept can be generalized to MPOs.
The coefficients $w_{\bm{\sigma \sigma'}}$ of a general operator
\begin{equation}
\widehat{\mathcal{W}} = \sum_{\bm{\sigma}, \bm{\sigma'}} w_{\bm{\sigma} \bm{\sigma'}}
    |\bm{\sigma} \rangle \langle \bm{\sigma'} |, 
    \label{eq:mpo}
\end{equation}
~\\
may be encoded in matrix-product form as
\begin{equation}
    w_{\bm{\sigma}, \bm{\sigma'}} =
    \sum_{b_1, \ldots, b_{L-1}}
    W^{\sigma_1 \sigma_1'}_{1 b_1}\, \cdots
    \,  W^{\sigma_l \sigma_l'}_{b_{l-1} b_l}\, \cdots
    \,  W^{\sigma_L \sigma_L'}_{b_{L-1} 1}.
\label{eq:mpo_coeff}
\end{equation}
Since we are mainly interested in operators corresponding to scalar observables, the two
indices $b_0$ and $b_L$ are restricted to $1$, such that we may later express the contraction
over the $b_i$ indices again as matrix-matrix multiplications yielding a scalar sum of operator terms.

Combining Eqs.\ \eqref{eq:mpo} and \eqref{eq:mpo_coeff}, the operator $\widehat{\mathcal{W}}$ reads
\begin{equation}
\widehat{\mathcal{W}} = \sum_{\bm{\sigma} \bm{\sigma'}}
    \sum_{b_1, \ldots, b_{L-1}} \!\!\!\!\!
    W^{\sigma_1 \sigma_1'}_{1 b_1}\, \cdots
    W^{\sigma_l \sigma_l'}_{b_{l-1} b_l} \cdots
    W^{\sigma_L \sigma_L'}_{b_{L-1} 1}  \,\,
    |\bm{\sigma} \rangle \langle \bm{\sigma'} |.
    \label{eq:mpo_tot}
\end{equation}
In contrast to the MPS tensors $M^{\sigma_l}$ in
Eq.\ \eqref{eq:mps}, the $W^{\sigma_l \sigma_l'}_{b_{l-1} b_l}$ tensors in Eq.\ \eqref{eq:mpo_tot}
each have an additional site index as superscript originating from the bra-ket notation in Eq.\ \eqref{eq:mpo}.
To simplify Eq.\ \eqref{eq:mpo_tot}, we perform a contraction
over the local site indices $\sigma_l, \sigma_l'$ in $\bm{\sigma}, \bm{\sigma'}$ by defining
\begin{equation}
\widehat{W}^l_{b_{l-1} b_l} = \sum_{\sigma_l, \sigma_l'} W^{\sigma_l \sigma_l'}_{b_{l-1} b_l}
    |\sigma_l\rangle \langle \sigma_l' |,
\end{equation}
so that Eq.\ \eqref{eq:mpo_tot} reads
\begin{equation}
\widehat{\mathcal{W}} =
    \sum_{b_1, \ldots, b_{L-1}} \!\!\!\!\!
    \widehat{W}^1_{1 b_1}\, \cdots
    \widehat{W}^l_{b_{l-1} b_l} \cdots
    \widehat{W}^L_{b_{L-1} 1}.
    \label{eq:mpo_thight}
\end{equation}
The motivation for this change in notation is that the entries of the resulting
$ \widehat{W}^l_{b_{l-1} b_l} $ matrices are the elementary operators acting on a single site,
such as the creation and annihilation
operators $\hat{c}^\dagger_{l \sigma}$ and $\hat{c}_{l \sigma}$.
To see this, note
that we may write, for instance,
\begin{equation}
\hat{c}^\dagger_{\uparrow} = 
 |\! \uparrow \! \downarrow \rangle \langle \downarrow \! | +
 | \! \uparrow \, \rangle \langle \,0\, |,
\end{equation}
which in practice can be represented
as a $4 \times 4$ matrix with two non-zero entries equal to $1$.
Hence, $ \widehat{W}^l_{b_{l-1} b_l} $ essentially collects all operators acting on site $l$
in matrix form. If we again recognize the summation over pairwise
matching indices $b_i$ as matrix-matrix
multiplications, we may drop them and rewrite Eq.\ \eqref{eq:mpo_thight} as
\begin{equation}
\widehat{\mathcal{W}} = \widehat{W}^1 \, \cdots \, \widehat{W}^L.
\label{eq:mpo_minimal}
\end{equation}

In practical applications one needs to find
compact representations of operators corresponding to
physical observables. For our purposes, the full electronic Hamiltonian,
\begin{equation}
  \widehat{\mathcal{H}} = \sum^L_{ij \, \sigma} t_{ij}
                    \hat{c}^\dagger_{i\sigma} \hat{c}^{\phantom{\dagger}}_{j\sigma}
                    + \frac{1}{2} \sum^L_{\substack{ijkl \\ \sigma \, \sigma'}} V_{ijkl}
                    \hat{c}^\dagger_{i\sigma} \hat{c}^\dagger_{k\sigma'}
                \hat{c}^{\phantom{\dagger}}_{l\sigma'} \hat{c}^{\phantom{\dagger}}_{j\sigma},
  \label{eq:hamil}
\end{equation}
is particularly important.
The MPO formulation now allows us to arrange the creation and annihilation operators
in Eq.\ \eqref{eq:hamil} into the operator valued matrices $\widehat{\mathcal{W}}$
from Eq.\ \eqref{eq:mpo_minimal}. In the following, we present two ways to find
such an arrangement. First, a very simple scheme is discussed to explain the basic concepts
with a simple example. Then, to obtain the same operational count as the DMRG in its traditional
formulation, we describe in a second step, how the $\widehat{\mathcal{W}}$ matrices
can be constructed in an efficient implementation.

\subsection{Na\"{\i}ve construction}

We start by encoding
the simplest term appearing in the electronic Hamiltonian of Eq.\ (\ref{eq:hamil}),
$t_{ij}  \hat{c}^\dagger_{i\uparrow} \hat{c}^{\phantom{\dagger}}_{j\uparrow}$,
as an MPO $\widehat{\mathcal{T}}(i,j)$.
With the identity operator $\hat{I}$ on sites $l \neq i,j$, we may write
\begin{equation}
  \widehat{\mathcal{T}}(i,j) =
    t_{ij} \cdot \hat{I}_1 \otimes \cdots \otimes \hat{c}^\dagger_{i\uparrow} \otimes \hat{I}_{i+1}
    \otimes \cdots \otimes \hat{c}_{j\uparrow} \otimes \cdots \otimes \hat{I}_L.
  \label{eq:opstring}
\end{equation}
For simplicity, we neglected the inclusion of fermionic anticommutation, which will be discussed in
section \ref{sec:fermions}.
To express this operator as an MPO, we need to find operator valued matrices
$\widehat{T}^i$, such that 
$\widehat{\mathcal{T}}(i,j) = \widehat{T}^1 \, \cdots \, \widehat{T}^L$.
In this case, there is only one elementary operator per site available. The only
possible choice is therefore
\begin{equation}
\widehat{T}^l_{11} = \hat{I}, \, l \neq i, j, \quad \widehat{T}^i_{11}
                   = t_{ij} \cdot \hat{c}^\dagger_{i\uparrow}, \quad
                \widehat{T}^j_{11} = \hat{c}^{\phantom{\dagger}}_{j\uparrow},
    \label{eq:encode1}
\end{equation}
which recovers $\widehat{\mathcal{T}}(i,j)$ after multiplication. 
Note that the coefficient $t_{ij}$ could have been multiplied into any site and its indices
are therefore not inherited to the $\widehat{T}^{l}_{b_{l-1} b_l}$ notation.
By substituting the identity operators in Eq.\ \eqref{eq:opstring}
with elementary creation or annihilation operators at two additional sites, we can express two-electron terms as
an MPO. While this changes the content of the $\widehat{T}$ matrices at those sites
sites, their shape is left invariant.

Turning to the MPO representation of the Hamiltonian in Eq.\ \eqref{eq:hamil}, we enumerate
its terms in arbitrary order, such that we may write
\begin{equation}
\widehat{\mathcal{H}} = \sum_n \widehat{\mathcal{U}}^n,
\end{equation}
where $\widehat{\mathcal{U}}^n$ is the $n$-th term of the Hamiltonian in the format of Eq.\ \eqref{eq:opstring}.
By introducing the site index $i$, we can refer to the elementary operator on site $i$ of
term $n$ by $\widehat{\mathcal{U}}^n_i$ and generalize Eq.\ \eqref{eq:encode1} to
\begin{equation}
\widehat{H}^1_{1n} = \widehat{\mathcal{U}}^n_1, \quad \widehat{H}^i_{nn} = \widehat{\mathcal{U}}^n_i, \, i \neq 1,L, \quad
                \widehat{H}^L_{n1} = \widehat{\mathcal{U}}^n_L.
  \label{eq:nterm}
\end{equation}
The formula above states how a sum of an arbitrary number of
terms can be encoded in matrix-product form. Since all matrices are diagonal,
except $\widehat{H}^1$ and $ \widehat{H}^L$ which are row and column vectors respectively,
the validity of
\begin{equation}
\widehat{\mathcal{H}} = \widehat{H}^1 \, \cdots \, \widehat{H}^L
\end{equation}
is simple to verify.

As the number of terms in the Hamiltonian operator is equal to the number of
non-zero elements of each $\widehat{W}^i$,
the cost of applying the MPO on a single site scales as
$\mathcal{O}(L^4)$ with respect to the number of sites,
and hence the total operation of $\widehat{\mathcal{H}} | \psi \rangle$
possesses a complexity of $\mathcal{O}(L^5)$. This simple scheme of
constructing the Hamiltonian operator in matrix-product form thus leads to an increase
in operational complexity by a factor of $L$ compared to the traditional formulation of the
DMRG algorithm for the quantum chemical Hamiltonian of Eq.\ \eqref{eq:hamil} \cite{White1999}.

\subsection{Compact construction}
It is possible to optimize the Hamiltonian MPO
construction and reduce the operational count by a factor of $L$.
We elaborate on the ideas presented in Refs. \cite{Crosswhite2008b} and \cite{Dolfi2014}, where identical subsequences among operator terms
in the form of Eq.\ \eqref{eq:opstring} are exploited.
If one considers tensor products as graphical connections between two sites, we can abstract
the term in Eq.\ \eqref{eq:opstring} to a single string running
through all sites.
Because of the nature of matrix-matrix multiplications,
each element in column $b$ of
$\widehat{W}^i$ will be multiplied by each element in row $b$ of $\widehat{W}^{i+1}$
in the product $\widehat{W}^i \cdot \widehat{W}^{i+1}$.
In the naive representation in Eq.\ \eqref{eq:nterm}
of the Hamiltonian in Eq.\ \eqref{eq:hamil},
operators were only placed on the diagonal, such that each matrix
entry on site $i$ is multiplied by exactly one entry on site $i+1$.
In a connectivity diagram, the construction discussed in the previous section
therefore results in parallel strings, one per term, with no interconnections between them.
%
 
\begin{figure}[H]
\includegraphics[width=\linewidth]{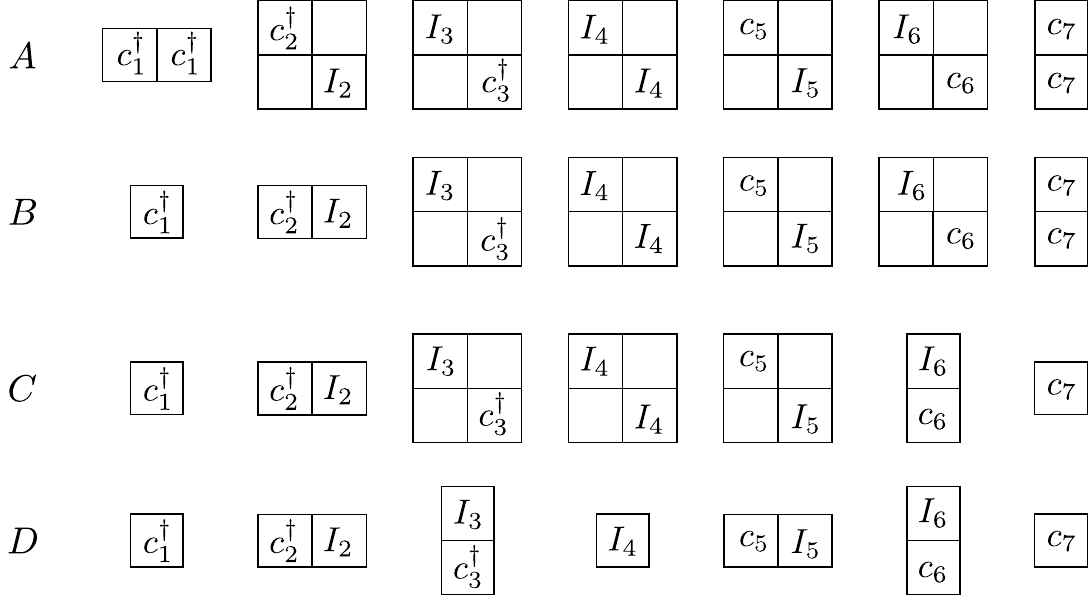}
\caption{The terms $c^\dagger_1 c^\dagger_2 c_5 c_7$ and $c^\dagger_1 c^\dagger_3 c_6 c_7$ are encoded
         in one MPO, whose matrices are depicted in the figure for the seven sites.
         The naive construction leads to the matrices shown in panel $A$. A \emph{fork} optimization
         was carried out in panel $B$ to exploit the common operator on site $1$.
         The combination of both \emph{fork} and \emph{merge} optimizations yields panel $C$. Finally,
         the attempt to exploit the common identity operator on site $4$ in panel $D$ by
         introducing an additional \emph{fork} and \emph{merge} optimization fails, as unwanted cross terms
         are generated. We note that $C$ is only possible if both terms have the same matrix elements ($1$ in this example).}
  \label{fig:MPOexample}
\end{figure}

%A reduction in the number of non-zero entries $\widehat{W}^i_{b_{i-1}, b_i}$
%and strings running in parallel on the bonds between sites
%can be achieved in two different ways.
The naive scheme may be improved in two different ways.
First, if two or more strings are identical
on sites $1$ through $k$, they can be collapsed
into one single substring up to site $k$, which is then \emph{forked} into new strings on site $k+1$,
splitting the common left part into the unique right remainders.
This graphical analogy translates into placing local operators on site $k+1$ on the
same row of the MPO matrix $\widehat{W}^{k+1}$.
Because each operator on the site of the fork
will be multiplied by the shared left part upon matrix-matrix multiplication, the
same operator terms are obtained.
The second option is to \emph{merge} strings that match on sites $l$ through $L$
into a common right substring, 
which is realized if local operators on site $l-1$ of the strings
to merge are placed on the same column in $\widehat{W}^{l-1}$.
If the Hamiltonian operator in Eq.\ \eqref{eq:hamil} is constructed in this
fashion, there will also be strings with identical subsections in the middle
of the lattice of sites. If we wanted to collapse, for instance, two of these shared substrings, we would first
have to merge strings at the start of the matching subsection and subsequently fork on the right end.
In general, however, it is not possible to collapse
shared substrings in this manner, because each of the two strings to the left of the merge site will
be multiplied by both forked strings on the right. Four terms are thus
obtained where there were initially only two.
As an example, the MPO manipulation techniques discussed
in this paragraph are illustrated in Fig.\ \ref{fig:MPOexample}.
We encoded the two terms
$c^\dagger_1 c^\dagger_2 c_5 c_7$
and
$c^\dagger_1 c^\dagger_3 c_6 c_7$
in an MPO according to the naive construction and subsequently applied \emph{fork} and \emph{merge} optimizations.

To compact the $\widehat{W}$ matrices,
we collapse strings from both sides of the orbital chain lattice
simultaneously.
Starting from the naive construction of the Hamiltonian operator,
each string is divided into substrings between the sites, which we call MPO bonds.
Next, we assign to each MPO bond a symbolic label which will
later indicate, where we may perform merges and forks.
The decisive information, that such a symbolic label carries,
is a list of (operator, position)-pairs which either have already
been applied to the left, implying forking behavior, or
which will be applied to the right, leading to merging behavior.
More specifically, if we assign an MPO bond between sites $k$ and $k+1$ with one of the labels
\begin{equation*}
    \text{fork:} \,\, \hat{c}^\dagger_{\uparrow},i ; \hat{c}_{\uparrow},j \quad \text{and} \quad
    \text{merge:} \,\, \hat{c}^\dagger_{\uparrow},l ; \hat{c}_{\uparrow},p \,\,,
\end{equation*}
we denote, in the first case that, on the current string, operators
$\hat{c}^\dagger_{\uparrow}$ and $\hat{c}_{\uparrow}$ were applied on sites $i$ and $j$
to the left and $i,j \leq k$.

We refer to this type of label as a \emph{fork}-label, because we defined
forking as the process of collapsing substrings to the left.
For this reason the label must serve as an identifier
for the left part of the current string.
In the second case, operators $\hat{c}^\dagger_{\uparrow}$ and $\hat{c}_{\uparrow}$
will be applied on sites $l$ and $p$ to the right, $l, p \geq k+1$, which serves as an identifier
for the right substring from site $k+1$ onwards.
For both types of labels we imply the presence of identity operators on sites not mentioned in the label.
Between each pair of sites, all MPO bonds with identical labels may now be collapsed into a single bond
to obtain a compact representation of the Hamiltonian operator.
Duplicate common substrings
between different terms are avoided in this way, because
the row and column labels will be identical within
shared sections.
In a final stage, at each MPO bond, we enumerate the symbolic labels,
in arbitrary order,
which yields the $b_{l-1}$ and $b_l$ matrix indices of $\widehat{W}^l_{b_{l-1} b_l}$.

\begin{figure}[H]
\includegraphics[width=\linewidth]{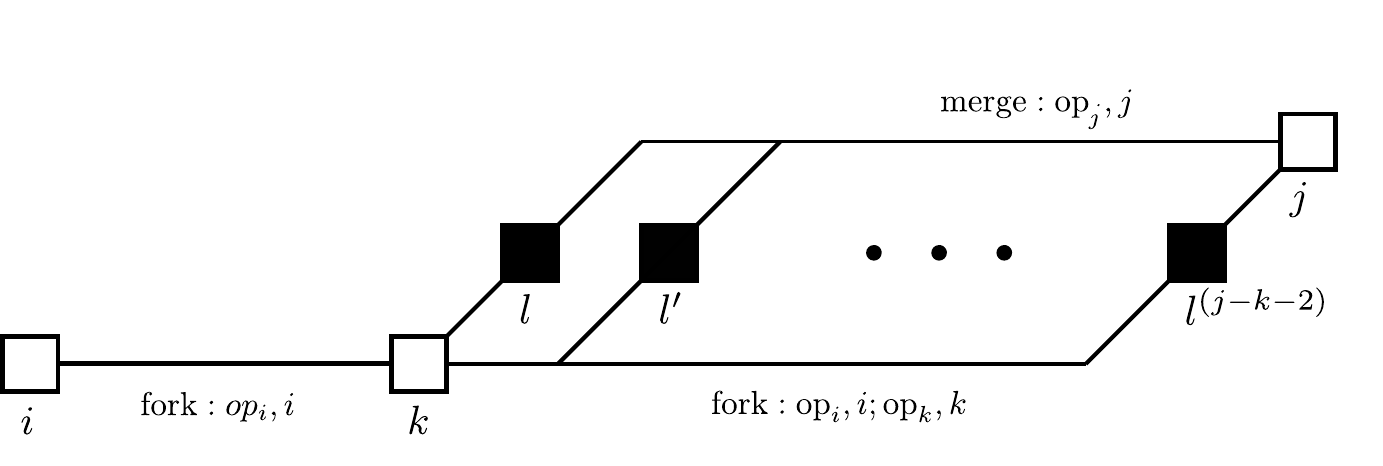}
\caption{The series of operators terms
    $V_{ijkl} \, \hat{c}^\dagger_{i,\sigma} c^\dagger_{k, \sigma'}
     \hat{c}_{l, \sigma'} \hat{c}_{j, \sigma}$, $l = k+1 \ldots j-1$, $i<k<l<j$,
    with combined common subsequences. Open squares correspond to operators,
    filled squares to operators multiplied by the matrix element $V_{ijkl}$.   }
  \label{fig:4term}
\end{figure}

This leads to the question of when to apply which type of label.
The general idea is to start with fork labels from the left and with merge
labels from the right.
Considering, for instance, a term
    $V_{ijkl} \, \hat{c}^\dagger_{i,\sigma} c^\dagger_{k, \sigma'}
     \hat{c}_{l, \sigma'} \hat{c}_{j, \sigma}$, {\small $i < k < l < j$}, %\normalsize,
the four non-trivial operators divide the string running over all sites into
the five substrings 1--$i$, $i$--$k$, $k$--$l$, $l$--$j$ and $j$--$L$
with different symbolic labels. Since we start with fork from the left and
merge from the right, we need to concern ourselves with the three substrings in the center.
Regarding the connectivity types of the labels for these substrings, the combinations
\emph{fork-fork-merge} or \emph{fork-merge-merge} lead to particularly
compact MPO representations.
The former combination is shown in Fig.\ \ref{fig:4term}:
all terms in the series
    $V_{ijkl} \, \hat{c}^\dagger_{i,\sigma} c^\dagger_{k, \sigma'}
     \hat{c}_{l, \sigma'} \hat{c}_{j, \sigma}$, {\small $l = k+1 \ldots j-1$, $i<k<l<j$} %\normalsize
share the operators at sites $i,k$ and $j$, such that the matrix element $V_{ijkl}$
has to be multiplied by the operator at site $l$, coinciding with the position at which the
label type changes from \emph{fork} to \emph{merge}.

\begin{figure}[H]
 \begin{center}
   \includegraphics[width=0.8\linewidth]{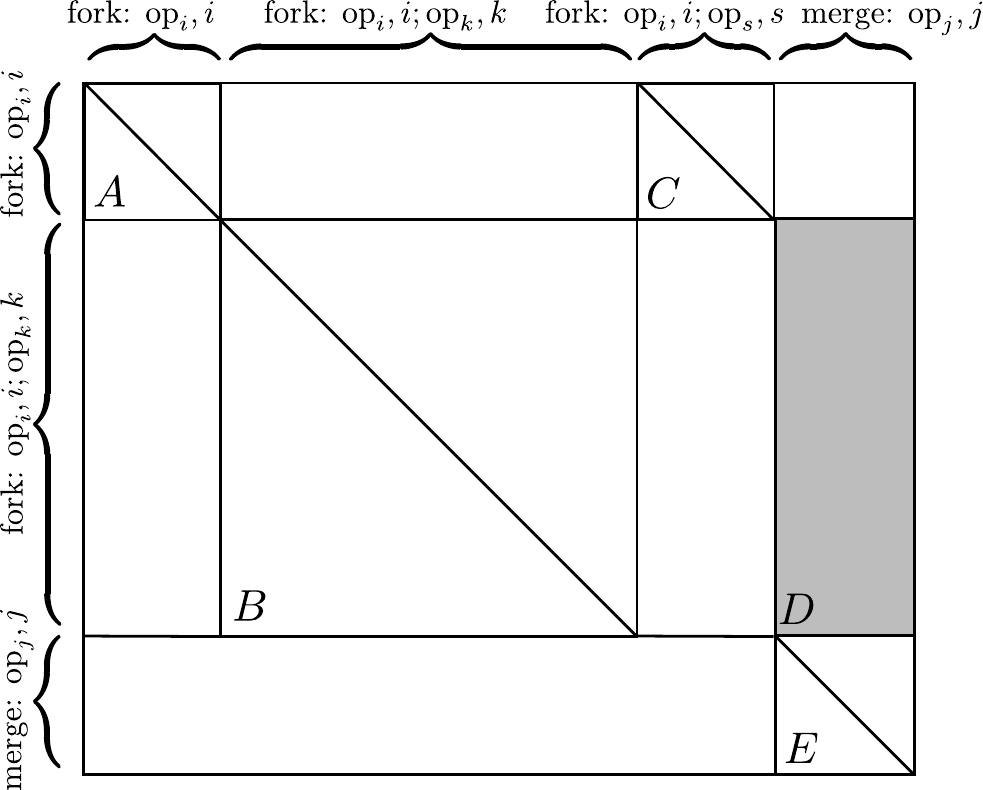}
 \end{center}
 %\vspace*{-1cm}
 \caption{MPO tensor $\widehat{W}^s_{b_{s-1} b_s}$ on some site $s$. $A$, $B$, $C$ and $E$ are diagonal
          blocks which extend incoming operator terms to the next site, while $D$ is dense and
          contains creation or annihilation operators multiplied by $V_{ijks}$.
          The first and last operator of each term, corresponding to $s=i$ and $s=j$, were omitted.
         }
 \label{fig:MPOscale}
\end{figure}

We now focus on one site $s$ only and describe the shape of
$\widehat{W}^s_{b_{s-1}b_s}$ for an MPO containing the terms
$\sum_{i \neq j\neq k \neq l} V_{ijkl} \, \hat{c}^\dagger_{i,\sigma} c^\dagger_{k, \sigma'}
     \hat{c}_{l, \sigma'} \hat{c}_{j, \sigma}.$
This simplification will retain the dominant structural features of the MPO for the
Hamiltonian in Eq.\ \eqref{eq:hamil}, because there are still
$\mathcal{O}(L^4)$ terms in this set.
Assigning all terms the described labels
and subsequent numbering on each MPO bond yields $\widehat{W}^s, \,s = 1,\ldots,L$. Fig.\ \ref{fig:MPOscale}
shows the result with labels sorted such that distinct subblocks can be recognized.
From their labels we can infer the following properties:\\

\begin{tabular}{lp{7.5cm}}
$A$ & One non-trivial operator ($\hat{c}^\dagger$ or $\hat{c}$) was applied
      to the left. \\
      %Fill operators $\hat{F}$ (see section \ref{sec:fermions}) continue
      %these terms to the next site.\\
$B$ & Continuation of terms with two non-trivial operators to the left
      (by means of identity operators).\\
$C$ & The application of the second non-trivial operator (if $k=s$) takes the same input as $A$
      but forks into labels with two operators, which will thus be part of the input to
      block $B$ and $D$ on the next site.\\
$D$ & If $l=s$, the third non-trivial operator is multiplied by $V_{ijkl}$ and placed
      in block $D$. There exists a term for every possible combination of operators pairs on sites $i,k$ 
      to the left and the operator on site $j$. Thus $D$ is dense.\\
$E$ & Terms with three non-trivial operators on the left are extended by $\hat{F}$ operators
      to site $s+1$ to eventually connect to the last non-trivial operator on site $j$.\\
\end{tabular}

Both rows and columns contain labels with at most two positions, implying that
the matrix size scales as $\mathcal{O}(L^2) \times \mathcal{O}(L^2)$.
The cost of contracting the operator with the MPS on one site is proportional to the
number of non-zero elements in the MPO matrix. As block $D$ has $\mathcal{O}(L^2)$ rows and $\mathcal{O}(L)$
columns, the algorithm will scale as $\mathcal{O}(L^4)$, because it performs $L$ %microiterations 
iterations
with a complexity of $\mathcal{O}(L^3)$ in one sweep.

Note that the first and last operator (i.e. $s=i$ and $s=j$) of each term is missing in Fig.\ \ref{fig:MPOscale}; their inclusion
only adds a constant number of rows and columns to $\widehat{W}^s$ (one for each of
$ \hat{c}^\dagger_{\uparrow}$, $c^\dagger_{\downarrow}$,
     $\hat{c}_{\uparrow}$, $\hat{c}_{\downarrow}$
).
Terms omitted from the previous discussion may easily be
accomodated in Fig.\ \ref{fig:MPOscale}.
The middle operator
of two-electron terms with one matching pair of indices, for instance, will enter the square above block $D$.

%%%%%%%%%%%%%%%%%%%%%%%%%%%%%%%%%%%%%%%%%%%%%%%%%%%%%%%%%%%%%%%%%%%%%%%%%%%%%%%%
\section{Fermionic anti-commutation}
\label{sec:fermions}
%%%%%%%%%%%%%%%%%%%%%%%%%%%%%%%%%%%%%%%%%%%%%%%%%%%%%%%%%%%%%%%%%%%%%%%%%%%%%%%%

In the operator construction scheme that we described in the previous section
we omitted the description of fermionic anticommutation.
\begin{align}
\lbrace \hat{c}^\dagger_{i \sigma}, \hat{c}_{j \sigma'} \rbrace& = \delta_{ij} \delta_{\sigma \sigma'}, \nonumber \\
\lbrace \hat{c}^\dagger_{i \sigma}, \hat{c}^\dagger_{j \sigma'} \rbrace = 
&\lbrace \hat{c}_{i \sigma}, \hat{c}_{j \sigma'} \rbrace = 0.
\label{eq:commut}
\end{align}
We can can include it by performing a
Jordan-Wigner transformation \cite{Jordan1928} on the elementary creation and annihilation operators.
Introducing the auxiliary operator
\begin{equation}
    \widehat{F} = \left( \begin{array}{c|cccc}
  & |\!\! \uparrow \! \downarrow \rangle & | \! \uparrow \, \rangle & | \! \downarrow \rangle & | \,0\, \rangle \\
\hline
          |\!\! \uparrow \! \downarrow \rangle & 1 & 0  & 0  & 0  \\
          |\! \uparrow \, \rangle              & 0 & -1 & 0 & 0 \\
          |\! \downarrow \, \rangle            & 0 & 0  & -1 & 0 \\
          |\, 0 \, \rangle                     & 0 & 0  & 0  & 1 
          \end{array} \right),
    \label{eq:fermions}
\end{equation}
we transform each elementary creation or annihilation operator
$\hat{\gamma}$ ($\hat{\gamma} = \hat{c}^\dagger_\sigma, \, \hat{c}_\sigma$) as
\begin{equation}
    \widehat{\gamma}_i \quad \mapsto \quad
      \Big( \prod_{s<i} \widehat{F}_s \Big ) \widehat{\gamma}_i
    \quad = \quad \widehat{\gamma}^{\mathrm{JW}}_i,
\label{eq:jwtrafo}
\end{equation}
such that Eqs.\ \eqref{eq:commut} are fulfilled by the transformed operators.
Since $\widehat{F}_s^2 = \mathbb{I}$ and $[\widehat{F}_s, \widehat{\gamma}_i] = 0$ if $s\neq j$
, we can simplify
\begin{equation}
  \widehat{\gamma}^{\mathrm{JW}}_i \widehat{\gamma}^{\mathrm{JW}}_j
    =
    \widehat{\mathbb{I}}_1 \otimes \ldots \otimes
    \widehat{\gamma}_i 
    \widehat{F}_i \otimes \widehat{F}_{i+1} \otimes \ldots \otimes \widehat{F}_{j-1} \otimes 
    \widehat{\gamma}_j,
    \otimes \ldots \otimes \widehat{\mathbb{I}}_L
\end{equation}
and find that in terms with an even number of fermionic operators, the sign operators 
only have to be applied in sections with an odd number of fermions to the left.

The described transformation ensures that fermions on different sites
anti-commute, but since the two kinds of electrons, $\uparrow$ and $\downarrow$, can coexist on the
same site, we have to modify Eq.\ \eqref{eq:jwtrafo} in order to obtain the correct behavior:
the operator $\hat{c}^\dagger_\downarrow$ has to be multiplied by $\widehat{F}$,
yielding $\hat{c}^\dagger_\downarrow \widehat{F}$ prior to the transformation in Eq.\ \eqref{eq:jwtrafo}
($\hat{c}_\downarrow \mapsto \widehat{F} \hat{c}_\downarrow$ by hermitian conjugation).
The additional sign operator keeps track of the phase factor incured by anti-commutation
among electrons on the same site with opposite spin.

%%%%%%%%%%%%%%%%%%%%%%%%%%%%%%%%%%%%%%%%%%%%%%%%%%%%%%%%%%%%%%%%%%%%%%%%%%%%%%%%
\section{MPS-MPO operations}
\label{sec:eval}
%%%%%%%%%%%%%%%%%%%%%%%%%%%%%%%%%%%%%%%%%%%%%%%%%%%%%%%%%%%%%%%%%%%%%%%%%%%%%%%%

Now that we have described the representation of many-electron states in the MPS format and
the representation of arbitrary operators as MPOs, we would like to calculate ground-states
of given Hamiltonians and subsequently compute expectation values on these states.

\subsection{Expectation Values}
\label{sec:expectation_values}
Based on the definitions in Eqs.\ \eqref{eq:mps} and \eqref{eq:mpo_thight}, 
we can immediately write down an expression for the transition operator matrix elements by
introducing a second state
\begin{equation}
    | \phi \rangle = \sum_{\bm{\sigma}}
        N^{\sigma_1} \, N^{\sigma_2} \, \cdots \, N^{\sigma_L}
        |\bm{\sigma} \rangle,
    \label{eq:mps2}
\end{equation}
or for expectation values if $|\phi \rangle = |\psi \rangle$:
\begin{align}
   \label{eq:expect_simple}
    \langle \phi | \widehat{\mathcal{W}} | \psi \rangle
     =   \sum_{\bm{\sigma\sigma'}}   \big(& N^{\sigma_1} \cdots  N^{\sigma_L} \big)^*  \!\!\!
         \sum_{b_1 \ldots b_{L-1}} \!\!\!\! W^{\sigma_1 \sigma_1'}_{1 b_1} \cdots
                            W^{\sigma_L \sigma_L'}_{b_{L-1} 1} \nonumber \\
                            \cdot \, \big(& M^{\sigma'_1}\cdots M^{\sigma'_L} \big),
\end{align}
denoting MPS tensors of $|\phi\rangle$ by $N$ and
assuming $\langle \bm{\sigma}|\bm{\sigma'} \rangle = \delta_{\bm{\sigma}\bm{\sigma'}}$.
This expression is exponentially expensive to evaluate, but suitable regrouping reduces the
operation count to $\mathcal{O}(L^4m^3)$:
\begin{align}
    \label{eq:expect_grouped}
    &\langle \phi | \widehat{\mathcal{W}} | \psi \rangle
     =  \sum_{\sigma_L\sigma'_L, b_{L-1}}
        N^{\sigma_L \dagger}
        W^{\sigma_L \sigma_L'}_{b_{L-1} 1}
        \Bigg(
        \cdots \sum_{\sigma_2 \sigma'_2, b_1} N^{\sigma_2 \dagger} W^{\sigma_2 \sigma_2'}_{b_1 b_2}            
     \nonumber\\
      & \qquad\qquad \cdot
        \Big(
        \sum_{\sigma_1\sigma'_1}
        N^{\sigma_1 \dagger}
        W^{\sigma_1 \sigma_1'}_{1 b_1}
        M^{\sigma'_1}
        \Big)
        M^{\sigma'_2}
        \cdots
        \Bigg)
        M^{\sigma'_L}. 
\end{align}
The iterative character of the previous expression is already apparent from the fact
that, proceeding from the innermost bracket outwards, at each step the tensor objects
of the respective site are added to the contracted expression in the center.
By introducing the initial value $\mathbb{L}^{0}_{b_0} = 1$, $b_0 = 1$ we may thus translate Eq.\
\eqref{eq:expect_grouped} into an iterative equation:
\begin{equation}
 \mathbb{L}^{l}_{b_l} = \sum_{\sigma_l \sigma'_l, b_{l-1}} N^{\sigma_l \dagger}
                W^{\sigma_l \sigma_l'}_{b_{l-1} b_l}
                \mathbb{L}^{l-1}_{b_{l-1}}
                M^{\sigma'_l}.
  \label{eq:move_boundary}
\end{equation}
Since $N^{\sigma_1 \dagger}$ and $M^{\sigma'_1}$ each only have one column and one row, respectively
(and $W^{\sigma_l \sigma_l'}_{b_{l-1} b_l}$ beeing a scalar), the dimensions of matrix multiplications match
at all steps. We refer to the objects $\mathbb{L}^l$ generated in this fashion as left boundaries,
whose structure may be understood as a vector indexed by $b_l$.
From Eqs.\ (\ref{eq:move_boundary}) and (\ref{eq:mps_long}) we
can further infer, that its elements are square matrices with indices $(a_l, \,a_l)$.
The last boundary $\mathbb{L}^L$, is a scalar as well as the first, and,
according to Eq.\ \eqref{eq:expect_grouped}, equal to the desired expectation value.

\begin{figure}[H]
 \begin{center}
   \includegraphics[width=0.8\linewidth]{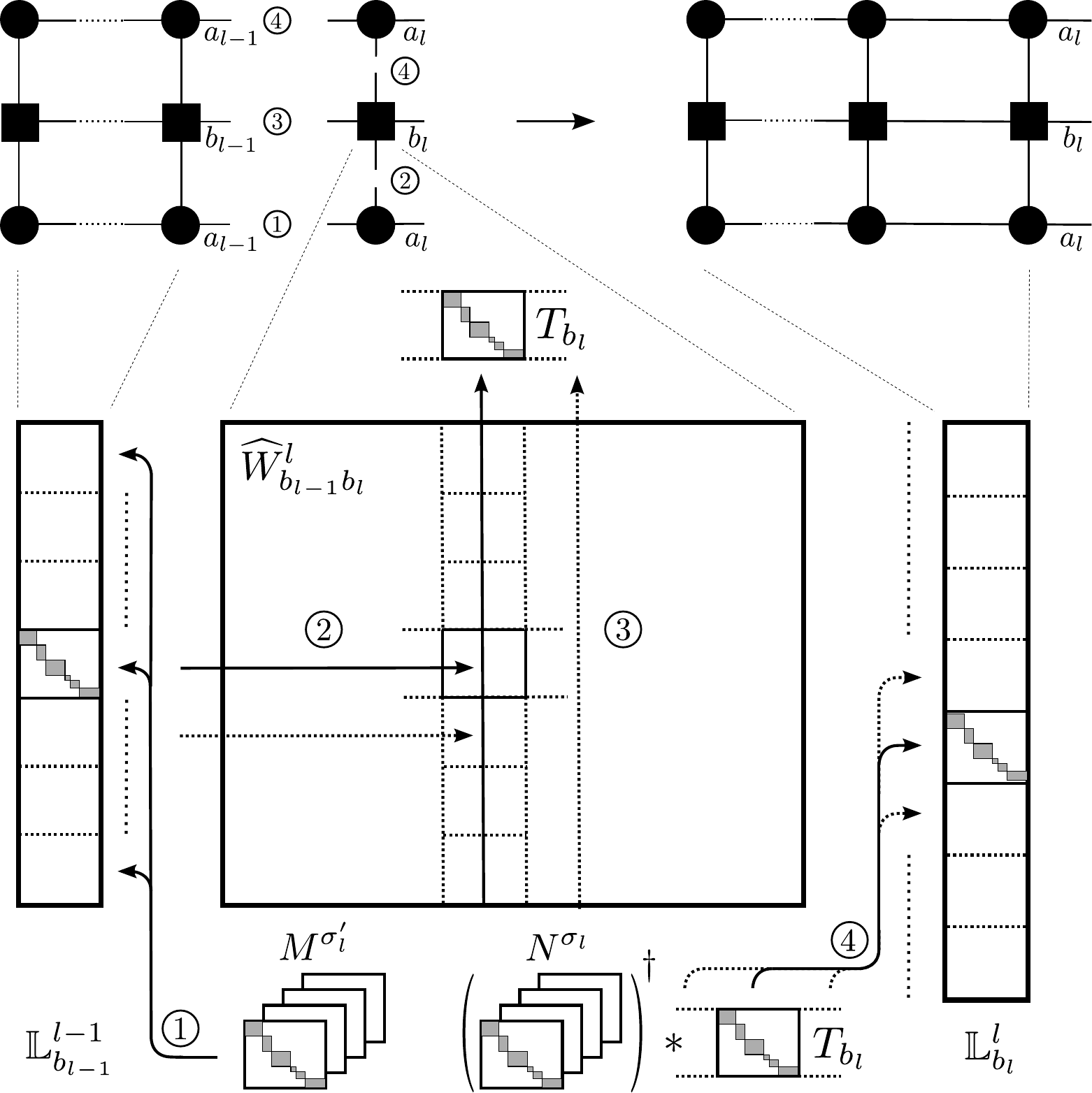}
 \end{center}
 %\vspace*{-1cm}
 \caption{Operations involved in enlarging the contracted left network $\mathbb{L}^{l-1}$ by one site
          to $\mathbb{L}^l$.
          Step 1: Contraction of $\mathbb{L}^{l-1}_{b_{l-1}}$ with the MPS tensor $M^{\sigma_l}$ (depicted
          as a set of four matrices, one for each value of $\sigma_l$) by matrix-matrix multiplication.
          Step 2: Summation over the local operator input $\sigma'_l$ of $W^{\sigma_l \sigma'_l}_{{b_{l-1} b_l}}$.
          Step 3: Summation over $b_{l-1}$. Step 4: Multiplication of $M^{\sigma_l \dagger}$ with the temporary
                  $T_{b_l}$ and summation over $\sigma_l$ yields $\mathbb{L}^l_{b_l}$.
          }
 \label{fig:MPOexpect}
\end{figure}

In Fig.\ \ref{fig:MPOexpect}, we illustrate the detailed operations contained in Eq.\ \eqref{eq:move_boundary}.
The schematic representation in the upper part abstracts MPS tensors $M^{\sigma_l}_{a_{l-1} a_l}$ as
circles and MPO tensors $W^{\sigma_l \sigma'_l}_{b_{l-1} b_l}$ as squares. Both diagrams
feature a leg for each tensor index sticking out 
and legs are connected to indicate a contraction over the two corresponding indices.
In detail, the boundary propagation starts with the contraction of $\mathbb{L}^{l-1}_{b_{l-1}}$ and
$M^{\sigma_l}$ by matrix-matrix multiplication (Fig.\ \ref{fig:MPOexpect} step 1), after which
the resulting products are multiplied by the MPO coefficient and have their quantum number mapped
from $\sigma'_l$ to $\sigma_l$ in step 2.
As the final product with $N^{\sigma_l \dagger}$ does not involve
$b_{l-1}$, the summation over $b_{l-1}$ into the temporary objects
\begin{equation}
    T_{b_l} = \sum_{\sigma'_l, b_{l-1}}
                W^{\sigma_l \sigma_l'}_{b_{l-1} b_l}
                \mathbb{L}^{l-1}_{b_{l-1}}
                M^{\sigma'_l}.
  \label{eq:t_temporary}
\end{equation}
is most efficiently performed
at this stage (step 3) and finally followed by multiplication with the bra MPS tensor $N^{\sigma_l\dagger}$
(step 4), yielding $\mathbb{L}^{l}_{b_l}$.

At this point we remark, that we could have started to contract the tensors from the right hand side as well,
which leads us to define right boundaries as
\begin{equation}
 \mathbb{R}^{l-1}_{b_{l-1}} = \sum_{\sigma_l \sigma'_l, b_{l-1}}
                M^{\sigma'_l}
                W^{\sigma_l \sigma_l'}_{b_{l-1} b_l}
                \mathbb{R}^{l}_{b_{l}}
                N^{\sigma_l \dagger}
                .
  \label{eq:move_boundary_right}
\end{equation}
From Eq.\ \eqref{eq:expect_grouped} we deduce that they can be assembled to yield the
desired expectation value with a matching left boundary as
\begin{equation}
    \langle \phi | \widehat{\mathcal{W}} | \psi \rangle = \sum_{b_l} \text{tr}\big(\mathbb{L}^l_{b_l} \mathbb{R}^l_{b_l} \big).
    \label{eq:expect_boundary}
\end{equation}
While these variants may seem redundant at this point, they will later allow us to maximize
the reuse of contracted objects during the sweeping procedure of the ground-state search.

\subsection{Ground-state search}

While in traditional DMRG, ground-states are determined by repeatedly diagonalizing a superblock
Hamiltonian formed at shifting lattice bipartitions, in second generation DMRG discussed here,
the energy is minimized through a variational search with the MPS tensor entries as parameters.
According to standard MPS-DMRG theory \cite{Schollwock2011, Oestlund1995}, both approaches eventually result in
the diagonalization of the same local Hamiltonian matrices, which is tackled by large sparse eigensolver
techniques, such as Jacobi-Davidson.
To clarify how both formulations are related, we first reconstruct the left and right block basis
states as well as the superblock Hamiltonian from traditional DMRG in an MPO/MPS framework.
Subsequently, we derive the same eigenvalue equation for the local site optimization from a variational viewpoint.

For a bipartition into $\mathcal{L}$ and $\mathcal{R}$ at site $l$, the left and right block basis states
are given by
\begin{align}
  |a_{l-1}\rangle_{\mathcal{L}} &= \sum_{\sigma_1, \ldots, \sigma_{l-1}} \big
                                 (M^{\sigma_1} \cdots M^{\sigma_{l-1}}\big)_{1,a_{l-1}}
                                 |\sigma_1, \ldots, \sigma_{l-1} \rangle \\
 |a_l\rangle_{\mathcal{R}} &= \sum_{\sigma_{l+1}, \ldots, \sigma_L}
                                 \big( M^{\sigma_{l+1}} \cdots M^{\sigma_L}\big)_{a_l,1}
                                  |\sigma_{l+1}, \ldots, \sigma_{L} \rangle,
\end{align}
such that we can write the total state as
\begin{equation}
|\psi \rangle = \sum_{a_{l-1} a_l, \sigma_l} M^{\sigma_l}_{a_{l-1}, a_l} |a_{l-1}\rangle_{\mathcal{L}}
                |\sigma_l\rangle
                |a_l\rangle_{\mathcal{R}}.
\end{equation}
The matrix elements of the superblock Hamiltonian
$\langle a_{l-1} \sigma_l a_l | \widehat {\mathcal{H}} | a'_{l-1} \sigma'_l a'_l\rangle$
can then be calculated
by substituting $|a_{l-1}\rangle_{\mathcal{L}}$, $|a_l\rangle_{\mathcal{R}}$ and $\widehat{\mathcal{H}}$
with the corresponding MPO expressions, which leads to the eigenvalue problem
\begin{equation}
 \varepsilon \, M^{\sigma_l}_{a_{l-1} a_l} =
  \sum_{a'_{l-1}a'_l,\sigma_l} \!\!\!\!\!
           \langle a_{l-1} \sigma_l a_l | \widehat {\mathcal{H}} | a'_{l-1} \sigma'_l a'_l\rangle
             \,\, M^{\sigma'_l}_{a'_{l-1} a'_l}.
\end{equation}
Alternatively, we can minimize the energy
$\langle \psi | \widehat {\mathcal{H}} | \psi \rangle$ with respect to the entries
of one MPS tensor $M^{\sigma_l}$ under the constraint that
$\langle \psi| \psi \rangle =1$ and arrive at the same eigenvalue equation \cite{Oestlund1995, Schollwock2011}.
Starting from Eq.\ \eqref{eq:expect_boundary}, we first expose the coefficients $M^{\sigma_l}_{a_{l-1} a_l}$ in
\begin{align}
    \langle \psi | \widehat{\mathcal{H}} | \psi \rangle = \sum_{b_l} \,\,\,&\text{tr}\big(\mathbb{L}^l_{b_l} \mathbb{R}^l_{b_l}\big) \\
    = \sum_{b_{l-1} b_l, \sigma_l \sigma'_l} \!\!\!\!\!\!
                     &\text{tr} \big(
                     M^{\sigma_l \dagger}
                     W^{\sigma_l \sigma'_l}_{b_{l-1} b_l}
                     \mathbb{L}^{l-1}_{b_{l-1}}
                     M^{\sigma'_l} \mathbb{R}^l_{b_l},
                     \big)
\end{align}
such that we can minimize the energy with respect to $M^{\sigma_l}_{a_{l-1} a_l}$ by
introducing a Lagrangian multiplier $\lambda$:
$\delta[\langle \psi | \widehat {\mathcal{H}} | \psi \rangle - \lambda (\langle \psi | \psi \rangle-1)] = 0.$
Under the assumption that $\mathbb{L}^l_{b_l}$ and $\mathbb{R}^l_{b_l}$ were formed from left and right normalized
MPS tensors respectively, the derivative of $\langle \psi | \psi \rangle$ with respect to
$M^{\sigma_l,\star}_{a_{l-1}a_l}$ is just $M^{\sigma_l}_{a_{l-1}a_l}$ and we arrive at the standard
eigenvalue problem
\begin{align}
     \sum_{b_{l-1} b_l, \sigma'_l} \!\!\!\!\!\!
                     W^{\sigma_l \sigma'_l}_{b_{l-1} b_l}
                     \mathbb{L}^{l-1}_{b_{l-1}}
                     M^{\sigma'_l} \mathbb{R}^l_{b_l} = \lambda M^{\sigma_l}.
    \label{eq:matrix_vector}
\end{align}
The computationally relevant operation is the matrix-vector multiplication
on the left hand side of the equation above as it is the key operation in sparse matrix eigensolvers.
Looking at this operation in detail, we find
that it is very similar to the boundary iteration in Eq.\ \eqref{eq:move_boundary}: steps 1--3
from Fig.\ \ref{fig:MPOexpect} are identical, but instead of multiplying the temporaries $T_{b_l}$
by the MPS conjugate tensor, a scalar product with $\mathbb{R}^l_{b_l}$ is formed.
With the implementation of Eqs.\ \eqref{eq:move_boundary}, \eqref{eq:move_boundary_right}, and
\eqref{eq:matrix_vector} we have thus all operations at hand to compute ground-states of
Hamiltonian MPOs. Moreover, in case of the quantum chemical Hamiltonian of Eq.\ \eqref{eq:hamil},
the compression technique described in this work ensures the optimal execution time scaling of 
$\mathcal{O}(m^3 L^3) + \mathcal{O}(m^2 L^4)$.

In addition to the eigenvalue problem in Eq.\ \eqref{eq:matrix_vector} for a single site,
the procedure described in Ref. \cite{White2005} improves convergence by introducing
a tiny ad-hoc noise term which helps to avoid local minima by re-shuffling the renormalized states.

%%%%%%%%%%%%%%%%%%%%%%%%%%%%%%%%%%%%%%%%%%%%%%%%%%%%%%%%%%%%%%%%%%%%%%%%%%%%%%%%
\subsection{Single-site and two-site DMRG}
%%%%%%%%%%%%%%%%%%%%%%%%%%%%%%%%%%%%%%%%%%%%%%%%%%%%%%%%%%%%%%%%%%%%%%%%%%%%%%%%
In the previous section, we described the optimization of a single site, which
in practice has a high probability to become trapped in a local energy minimum.
Although the convergence behaviour may be improved by introducing
a tiny ad-hoc noise term \cite{White2005}, 
for chemical applications,
the so-called two-site DMRG algorithm, in which two sites
are optimized at the same time, achieves faster convergence and is much less likely
to get stuck in local minima.
In the MPS-MPO formalism, we can optimize two sites
at once by introducing the two-site MPS tensor
\begin{equation}
P^{\sigma_l \sigma_{l+1}}_{a_{l-1} a_{l+1}} = \sum_{a_l} M^{\sigma_l}_{a_{l-1} a_l} M^{\sigma_{l+1}}_{a_l a_{l+1}}
\end{equation}
and the two-site MPO tensor
\begin{equation}
\widehat{V}^{\sigma_l \sigma_{l+1} \sigma'_l \sigma'_{l+1}}_{b_{l-1} b_{l+1}}
    = \sum_{b_l} \widehat{W}^{\sigma_l \sigma_l'}_{b_{l-1} b_l} \widehat{W}^{\sigma_{l+1}\sigma'_{l+1}}_{b_l b_{l+1}}.
\end{equation}
The latter case corresponds to a matrix-matrix multiplication of the operator valued $\widehat{W}$-
matrices, whose entries are multiplied by forming the tensor product of the local site operators.
If $\sigma_{l}\sigma_{l+1} = \sigma_l \otimes \sigma_{l+1}$ is treated as a single 16-dimensional local space
$\tau_{l,l+1}$,
we can extend Eq.\ \eqref{eq:matrix_vector} from the previous section to two-site DMRG by exchanging
$M$ with $P$ and $W$ with $V$ and obtain
\begin{align}
     \sum_{b_{l-1} b_{l+1}, \tau'_{l,l+1}} \!\!\!\!\!\!
                     V^{\tau_{l,l+1} \tau'_{l,l+1}}_{b_{l-1} b_{l+1}}
                     \mathbb{L}^{l-1}_{b_{l-1}}
                     P^{\tau'_{l,l+1}} \mathbb{R}^{l+1}_{b_{l+1}} = \lambda P^{\tau_{l,l+1}}.
    \label{eq:matrix_vector_twosite}
\end{align}
The different steps in evaluating the above expression are essentially the same
as for the single-site case, such that the same program routines can be used
for both optimization schemes after the generation of the two-site tensors.

\subsection{Excited states}

The MPS-based state specific algorithm to economically calculate excited states \cite{McCulloch2007}
repeatedly orthogonalizes the excited state against a supplied list
of orthogonal states at every micro-iteration during the Jacobi-Davidson diagonalization step.
Let $|\phi_n\rangle$ denote the $n$-th supplied orthogonal state and $|\psi\rangle$ the target wave function.
We can then define the partial overlaps
\begin{equation}
    C^l_n = \sum_{\sigma_l} M^{\sigma_l \dagger} \cdots \Big(\sum_{\sigma_1} M^{\sigma_1 \dagger} N^{\sigma_1}_n \Big) \cdots N^{\sigma_l}_n
\end{equation}
and
\begin{equation}
    D^{l-1}_n = \sum_{\sigma_l} N^{\sigma_l}_n \cdots \Big( \sum_{\sigma_L} N^{\sigma_L}_n M^{\sigma_L \dagger} \Big) \cdots M^{\sigma_l \dagger},
\end{equation}
where $|\phi_n\rangle$ and $|\psi\rangle$ are defined according to Eqs.\ \eqref{eq:mps} and \eqref{eq:mps2},
yielding $\langle \psi | \phi_n \rangle = \text{tr}(C^l_n D^l_n)$ for the overlap.

At every site $l$, the orthogonal vectors $V_n$, which the Jacobi-Davidson eigensolver takes as input parameters, can now be calculated as
\begin{equation}
    V_n = \sum_{\sigma_l} C^{l-1}_n N^{\sigma_l}_n D^{l}_n.
\end{equation}
We note, that $V_n$ has the property $\text{tr}(M^{\sigma_l \dagger}V_n) = \langle \psi | \phi_n \rangle$.
After diagonalization, $C^{l}_n$ and $D^{l-1}_n$ are updated with the optimized $M^{\sigma_l}$ tensor during a left and right sweep respectively.
Finally, the converged $|\psi\rangle$ will be orthogonal to all $|\phi_n\rangle$.

%%%%%%%%%%%%%%%%%%%%%%%%%%%%%%%%%%%%%%%%%%%%%%%%%%%%%%%%%%%%%%%%%%%%%%%%%%%%%%%%
\section{Calculations}
%%%%%%%%%%%%%%%%%%%%%%%%%%%%%%%%%%%%%%%%%%%%%%%%%%%%%%%%%%%%%%%%%%%%%%%%%%%%%%%%
\label{sec:calculations}
\begin{figure}[h]
  \includegraphics[width=0.9\linewidth]{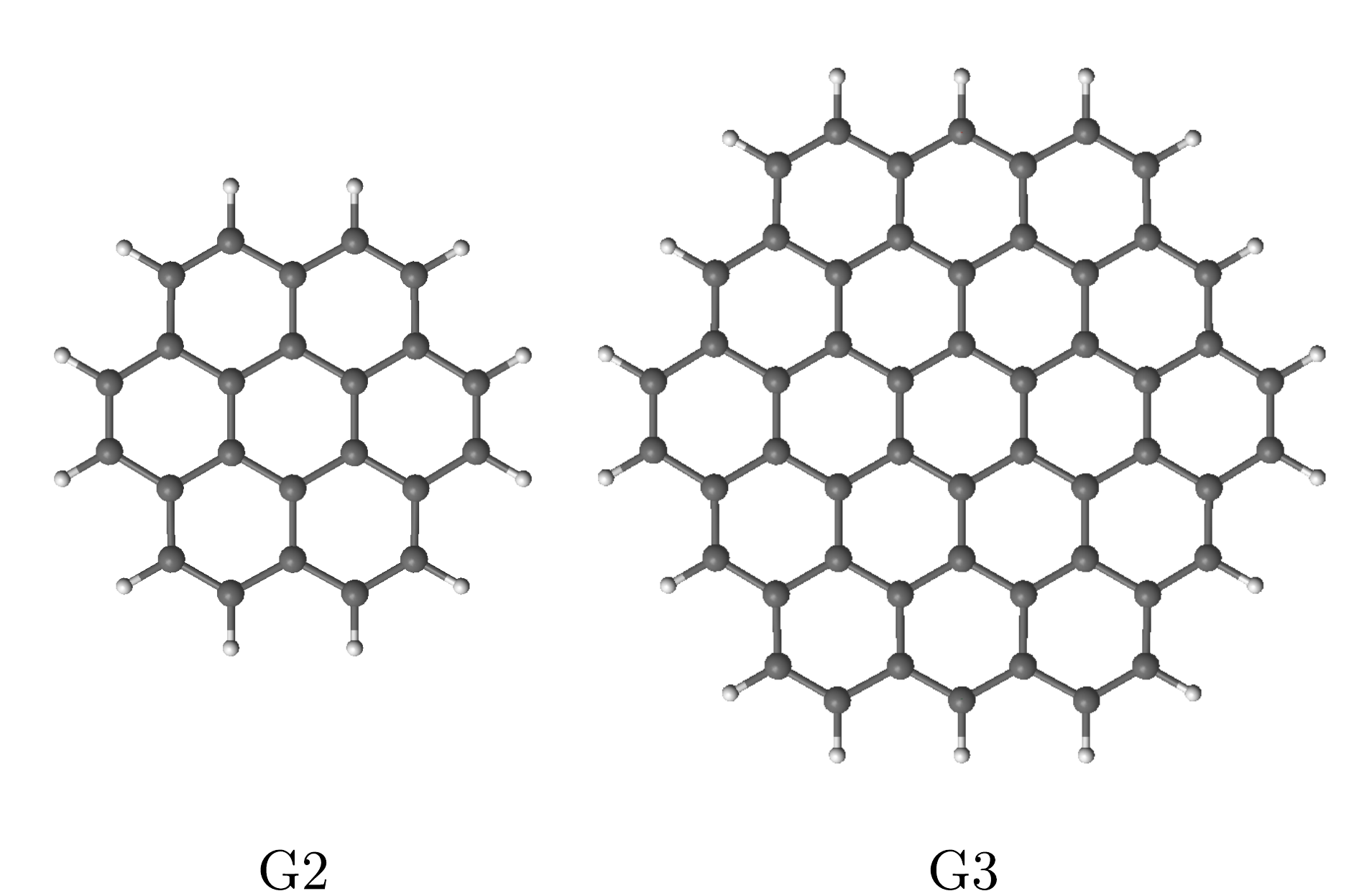}
  \caption{Graphene fragments G2 (left) and G3 (right) with 24 and 54 $\pi$-orbitals, respectively.}
  \label{fig:graphene}
\end{figure}
The efficiency of an MPO DMRG implementation critically depends on
the representation of the MPO.
As we have shown, both a compact construction and the exploitation
of sparsity are necessary to achieve optimal scaling.

To demonstrate its capabilities, we calculated
ground- and excited-state energies of the graphene fragments G2 and G3 shown in Fig.\ \ref{fig:graphene}.
The conjugation in both molecules provides for long-range correlation in two dimensions and they are therefore
difficult to treat with DMRG,
which performs best for one-dimensional and quasi one-dimensional systems.
In other systems with less delocalization and mixed active spaces, containing both $\sigma$- and $\pi$-type molecular orbitals
and possibly exhibiting spatial point group symmetry,
often a significant fraction of the two-electron integrals have negligible magnitude, leading to smaller MPOs such that
even larger active spaces become affordable.

\begin{figure}[t!]
  \includegraphics[width=0.9\linewidth]{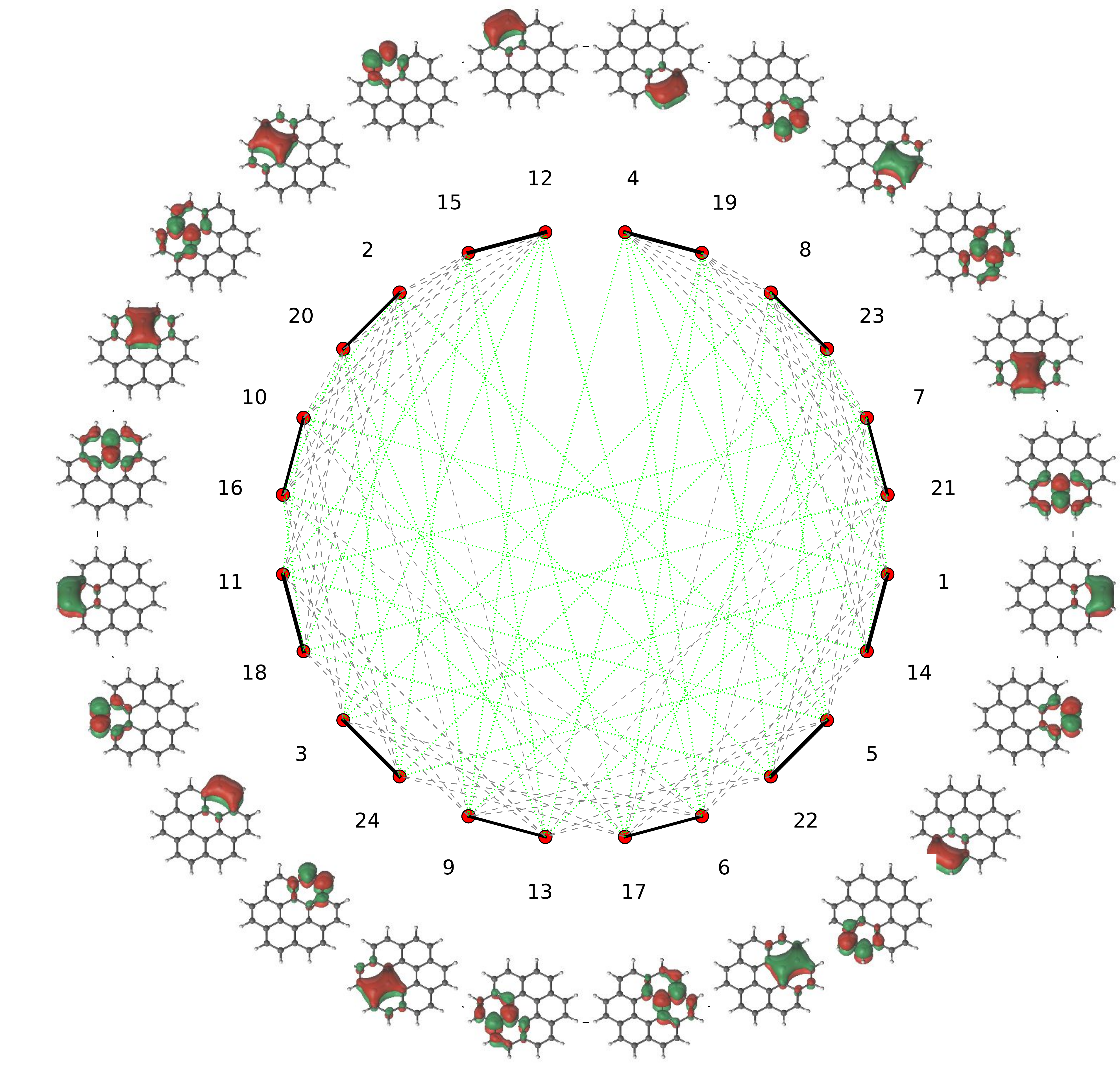}
  \caption{Orbital ordering for the fragment G2. The mutal information between pairs of orbitals is drawn with bold black lines
            for magnitudes $>0.1$, dashed gray lines ($0.01 - 0.1$) and dotted green lines $(0.001 - 0.1)$.}
  \label{fig:mutplot}
\end{figure}

We chose the $\pi$-orbitals for the active space, represented by a minimal STO-3G basis set \cite{Hehre1969}.
While this choice of basis precludes obtaining realistic energies, it does not simplify the
correlation problem within the active space, which we are interested to solve with DMRG.
A set of different condensed polyaromatic systems has recently been calculated by Olivares-Amaya et al. in Ref.\ \cite{Chan2015}, employing
the same basis set.
In accordance with their work, we found that performing Pipek-Mezey localization \cite{Pipek1989} first
within the active occupied orbitals, followed by a second Pipek-Mezey localization of
the active virtual orbitals, provided the molecular orbital basis with the best convergence.
The efficiency gain compared to Hartree-Fock orbitals more than compensated the loss of spatial
point group symmetry.

\begin{figure}[h]
  \includegraphics[width=\linewidth]{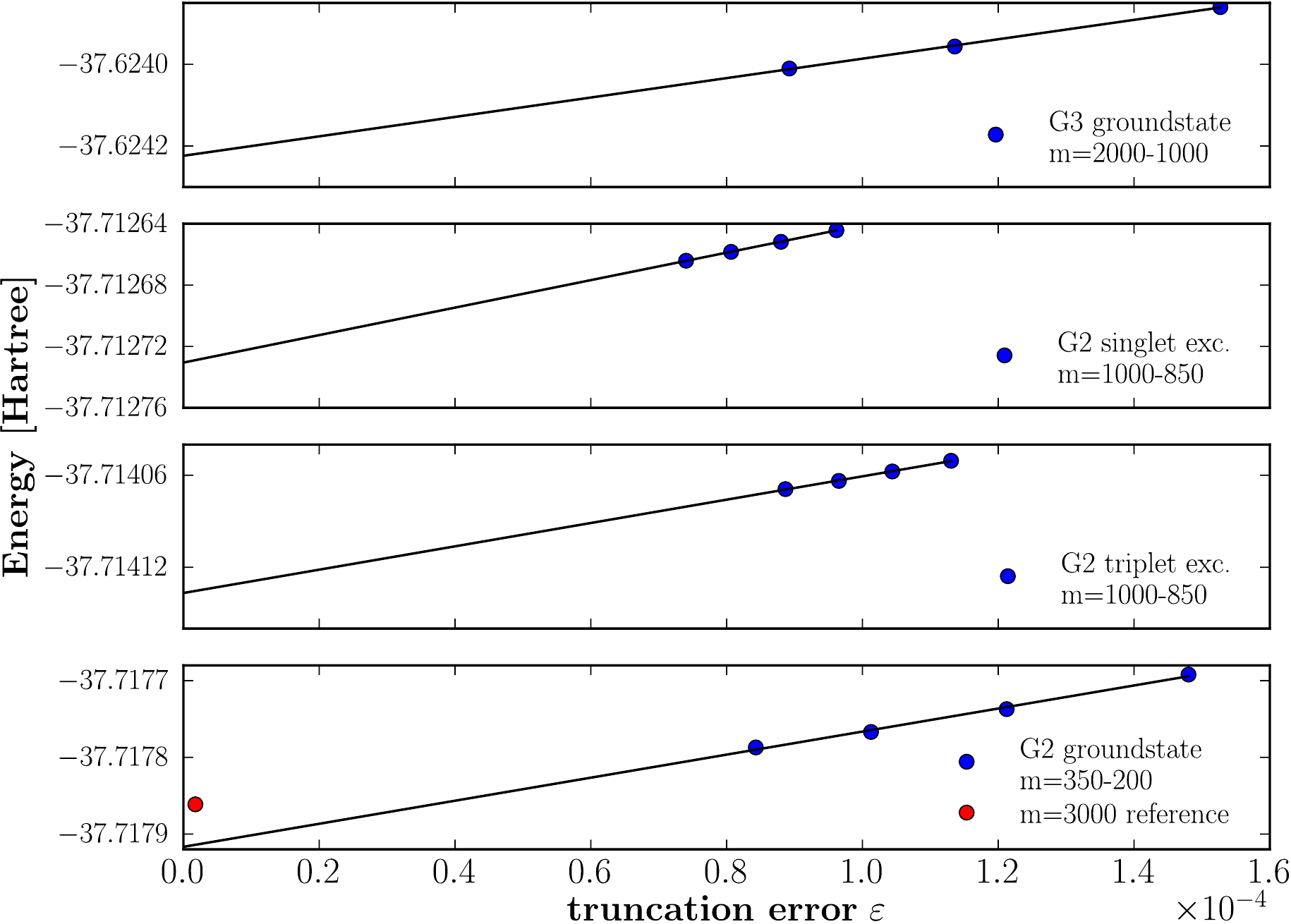}
  \caption{Energy per carbon atom in G2 and G3, extrapolated towards truncation error $\varepsilon \rightarrow 0$.
           First panel: ground-state of G3 with m = 2000, 1500 and 1000.
           Second panel: lowest singlet excited-state of G2 with m = 1000, 950, 900, 850.
           Third panel: lowest triplet excited-state of G2 with m = 1000, 950, 900, 850.
           Fourth panel: ground-state of G2 with m = 350, 300, 250, 200 and a reference calculation with m = 3000.}
  \label{fig:extrapolation}
\end{figure}

In order to obtain accurate energies, we exploit non-abelian spin symmetry whose implementation with the
quantum chemical DMRG Hamiltonian will be described in a later work \cite{SU2}.
Furthermore, the type of orbitals and the orbital ordering have to be carefully chosen to improve performance.
By computing the Fiedler vector \cite{Fiedler1973, Fiedler1975, Barcza2011}
based on the mutual information \cite{Rissler2006, Legeza2006}, we determined an efficient orbital ordering.
In Fig.\ \ref{fig:mutplot}, it is shown for the G2 fragment together
with the mutual information drawn as lines with varying thickness according to entanglement strength.
We observe that, based on this entanglement measure, the Fiedler ordering reliably grouped
strongly entangled bonding and anti-bonding orbitals pairs next to each other and arranged orbitals according to their
spatial position, such that the hexagonal fragment is traversed through the longest possible pathway across two adjacent corners.
The application of the Fiedler ordering to G3 led a similar type of orbital arrangment.

With this configuration, we performed ground- and excited-state calculations with the implementation reported in this paper.
The obtained energies per carbon atom are shown in Fig.\ \ref{fig:extrapolation}.
For the G2 ground-state calculation, $m$ was chosen such that we obtained truncation errors in the same
range as the G3 ground-state calculation.
This choice affords wavefunctions of similar quality in both cases and accordingly similar distances in the
extrapolated energy, with the advantage, that for the smaller G2 fragment we can verify the extrapolation error
with a precise reference calculation with $m = 3000$.
From the extrapolation, we estimate an error per carbon atom of
$\pm 0.05 \, \text{mH}$ for G2 and $\pm 0.1 \, \text{mH}$ for G3, respectively.
%
%The $m=2000$ calculation for G3 was converged after 10 sweeps, each of which required 27 hours on one 16-core node
%with Intel Xeon E5-2667v2 CPUs.

\newpage
\section{Conclusions}
\label{sec:conclusion}
\noindent
In this work, we provided a prescription for the efficient construction of the quantum chemical Hamiltonian in Eq.\ \eqref{eq:hamil}
as an MPO.
Our construction scheme ensures the same computational scaling as traditional DMRG, compared to which a
similar performance can be achieved.
The advantages of a full matrix product formulation of all wave functions and operators involved in the DMRG algorithm
are the efficient calculation of low-lying excited states, straightforward implementation of new observables, and additional
flexibility that allows us to accommodate different models and symmetries in a single source code.
We investigated our MPO approach at the example of two graphene fragments of different sizes for which we calculated ground and excited
state energies.
%Even though these systems are challenging to treat with DMRG due to delocalization in two dimensions, we achieved
%an accuracy of less than $1$ mH.

%\newpage
\section*{Acknowledgments}
This work was supported by ETH Research Grant ETH-34 12-2.

%\bibliographystyle{jcp}
%\bibliography{refs.bib}

\end{document}